\documentclass[%
preprint,
 amsmath,amssymb,
 aps,
 prl,
 superscriptaddress,
]{revtex4-1}

\usepackage{natbib}
\usepackage{amsmath, amssymb}
\usepackage{fancyhdr, ifpdf}
\usepackage{bm}
\usepackage{cases}

\usepackage{braket}
\usepackage{graphicx}
\usepackage{dcolumn}
\usepackage{bm}

\begin{document}

\title{Spontaneous two photon emission from a single quantum dot}

\author{Yasutomo Ota}
\email{ota@iis.u-tokyo.ac.jp}
\affiliation{Institute for Nano Quantum Information Electronics, The University of Tokyo, 4-6-1 Komaba, Meguro-ku, Tokyo 153-8505, Japan}%

\author{Satoshi Iwamoto}%
\affiliation{Institute for Nano Quantum Information Electronics, The University of Tokyo, 4-6-1 Komaba, Meguro-ku, Tokyo 153-8505, Japan}%
\affiliation{Institute of Industrial Science, The University of Tokyo, 4-6-1 Komaba, Meguro-ku, Tokyo 153-8505, Japan}%

\author{Naoto Kumagai}%
\affiliation{Institute for Nano Quantum Information Electronics, The University of Tokyo, 4-6-1 Komaba, Meguro-ku, Tokyo 153-8505, Japan}%

\author{Yasuhiko Arakawa}%
\affiliation{Institute for Nano Quantum Information Electronics, The University of Tokyo, 4-6-1 Komaba, Meguro-ku, Tokyo 153-8505, Japan}%
\affiliation{Institute of Industrial Science, The University of Tokyo, 4-6-1 Komaba, Meguro-ku, Tokyo 153-8505, Japan}%


\date{\today}

\begin{abstract}

Spontaneous two photon emission from a solid-state single quantum emitter is observed. We investigated photoluminescence from the neutral biexciton in a single semiconductor quantum dot coupled with a high Q photonic crystal nanocavity. When the cavity is resonant to the half energy of the biexciton, the strong vacuum field in the cavity inspires the biexciton to simultaneously emit two photons into the mode, resulting in clear emission enhancement of the mode. Meanwhile, suppression was observed of other single photon emission from the biexciton, as the two photon emission process becomes faster than the others at the resonance. 

\begin{description}
\item[PACS numbers] 42.50.Pq, 42.50.-p, 78.67.Hc, 42.70.Qs
\end{description}

\end{abstract}



\maketitle


Spontaneous two photon emission (TPE) is a quantum optical process, in which a pair of photons is simultaneously emitted, driven by vacuum field fluctuations. The process is thereby inherently weak compared to other single photon processes, but has importance in various research fields including astrophysics, atom physics~\cite{Breit1940,*Shapiro1959,*[{For a review paper, }] Ilakovac2006} and nonlinear optics~\cite{Burnham1970}. The generated two photons are useful in many applications, especially in quantum information technologies, as sources of heralded single photons~\cite{Rochester1985}, entangled photon pairs~\cite{Kwiat1995} and biphotonic qutrits~\cite{Bogdanov2004}. To these end, spontaneous parametric down conversion in nonlinear optical crystals~\cite{Burnham1970} is most commonly used, and recently, spontaneous TPE in highly-pumped semiconductors has recently emerged as a potential alternative~\cite{Hayat2008}. However, those systems produce the two photons intrinsically randomly with low efficiency. To prepare the two photons in a regulated manner~\cite{Bertet2002}, a straightforward choice is the utilization of single quantum emitters, such as gaseous atoms. 

TPE in such single quantum emitters have been investigated for decades~\cite{Lipeles1965}. To overcome competing single photon processes and efficient generation of the two photons, cavity effects have been often employed, which resulted in the realizations of two photon masers~\cite{Brune1987} and lasers~\cite{Nikolaus1981,Gauthier1992}. For those demonstrations, stimulated emission process of the two photons plays a dominant role and, so far, the physics in cavity-enhanced \textit{spontaneous} TPE has not been investigated in detail~\cite{[{A publication, } ][{, reported spontaneous TPE in a Rydberg atom in a cavity, however, the observed two photon processes are significantly affected by stimulation from thermal photons inside the cavity. }]Lange1996, *[{See also }]Enaki1999}. Moreover, the TPE process has not been investigated in single quantum emitters in the solid state, which have the potential to form robust platform for practical applications. 

In this letter, we demonstrate spontaneous TPE from a single quantum emitter in the solid state. A biexcitonic state in a single semiconductor quantum dot (QD) is studied under the strong coupling regime with a photonic crystal (PhC) nanocavity, the strong field confinement of which boosts the spontaneous TPE process. When a single cavity mode was tuned to the half energy of the biexciton, clear emission enhancement from the mode was observed. Simultaneously, we observed suppression of single photon cascaded emission from the biexciton to other single exciton states. Both these effects are consequences of the two photon resonance of the QD's biexcitonic state with the cavity mode, where the TPE process becomes faster than other possible single photon processes for the biexciton. The estimated cavity photon number is much below one, thus the observed TPE is considered to be spontaneous and driven predominantly by the enhanced vacuum field. The measured spectra are well reproduced by calculations based on a master equation including the two photon contribution.


We investigated a PhC double heterostructure cavity~\cite{Song2005} made from GaAs with the lattice constant $a$=250 nm, radius $r$ = 72.5 nm. $a$ for the double hetero region is elongated to 258 nm along the waveguide. The PhC slab has the thickness $d$ of 130 nm and a layer of InAs QDs with the density of $\sim$1.0~10$^{8}$cm$^{2}$ is buried at its center. The sample was kept at 4.5 K and investigated by a $\mu$-PL measurement setup. The excitation source is a continuous wave Ti:Sapphire laser oscillating at 780 nm and is focused onto the sample surface by a objective lens with a numerical aperture of 0.65. Throughout the measurement, the excitation power is kept at 180 nW  (measured before the objective lens). Collected PL signals by the same objective were sent to a spectrometer with the resolution of  23 $\mu$eV after passing though a polarizer parallel to the cavity polarization.

\begin{figure}
\centering
\includegraphics[width=\linewidth]{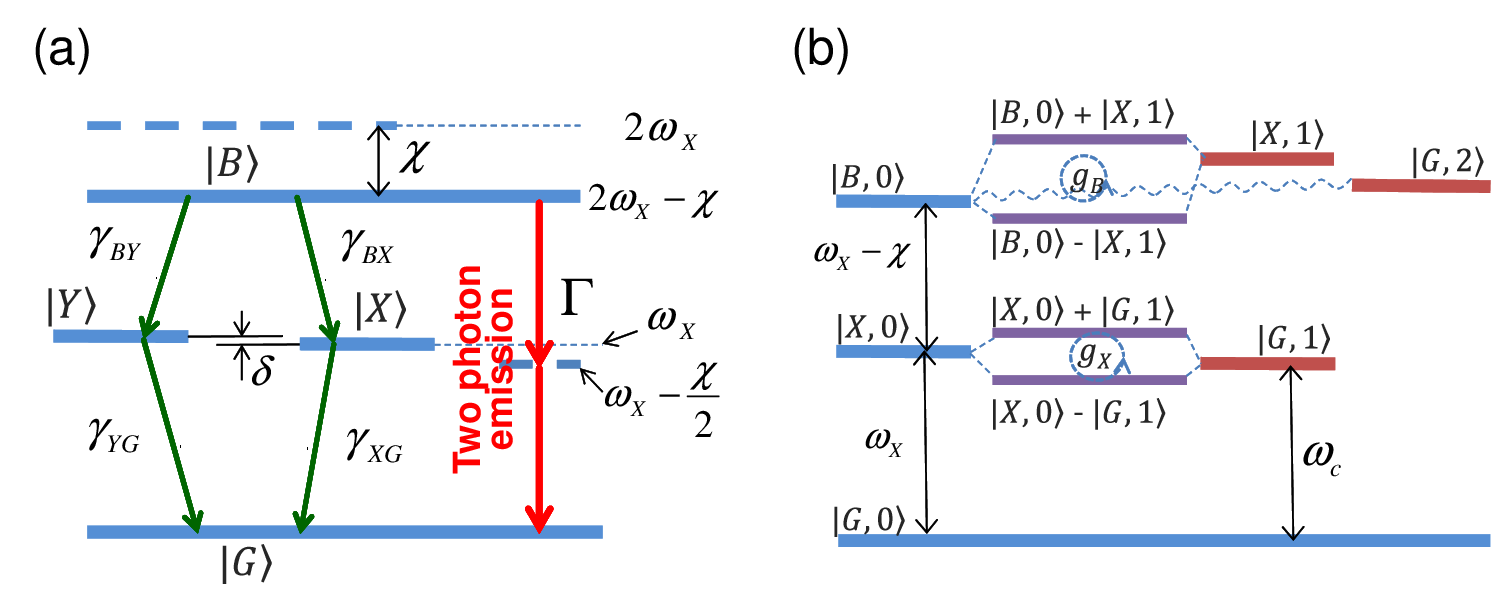}
\caption{(Color online) (a) Schematic illustration of the energy diagram of QD, including two single exciton states $\ket{X}$, $\ket{Y}$ and a biexciton state$\ket{B}$. (b) Relevant quantum states in the QD-cavity system for discussing the observed TPE. Weight of the superposition states, which significantly changes depending on the cavity detuning, is omitted for simplicity.}
\label{fig:f1}
\end{figure}

Figure ~\ref{fig:f1}(a) shows a energy level diagram of a QD described by 4 levels, some transitions of which are coupled with a single cavity mode at the frequency of $\omega_{c}$. The QD states are composed of a biexciton state $\ket{B}$, two single exciton states with orthogonal polarizations $\ket{Y}$, $\ket{X}$ and ground state $\ket{G}$, eigenfrequencies of which are $2\omega_{X}-\chi$, $\omega_{X}+\delta$, $\omega_{X}$ and 0, respectively. $\chi$ and $\delta$ express the biexciton binding energy and the fine structure splitting of the exciton states, respectively. The green arrows indicates dipole arrowed transitions in the QD. We assume that the linearly polarized cavity mode couples only with the transitions through $\ket{X}$. We denote spontaneous single photon emission rate between dipole arrowed levels $\ket{i} \rightarrow \ket{j}$ as $\gamma_{i,j}$ and the TPE rate as $\Gamma$. $\Gamma$ can be enhanced by using the two photon resonance with the cavity mode at $\omega_{c} = \omega_{X} - \chi /2$.

Energy levels related to QD-cavity coupling experiments in this letter are schematically illustrated in Fig.~\ref{fig:f1}(b). The single exciton state $\ket{X,0}$ couples with the single cavity photon state $\ket{G,1}$ with a rate $g_{X}$ under the resonant condition ($\omega_{c}$ = $\omega_{X}$). Through this coherent coupling, the states form polaritons, the features of which become prominent near the resonance. The biexciton state $\ket{B,0}$ couples with the state $\ket{X,1}$ by exchanging single photons and also with $\ket{G,2}$ by two photon transition with the approximate coupling strength of $4g^2/\chi$~\cite{[{}][{. To derive the effective coupling strength, large $\chi$ and $g_{X} \sim  g_{B} \sim g$ is assumed.}]DelValle2010} under the two photon resonant condition ($\omega_{c}$ = $\omega_{X} -\chi/2$). In this system, various optical transitions among the bare QD states, cavity, and polaritons are observable depending on the cavity detuning.

\begin{figure}
\centering
\includegraphics[width=7cm]{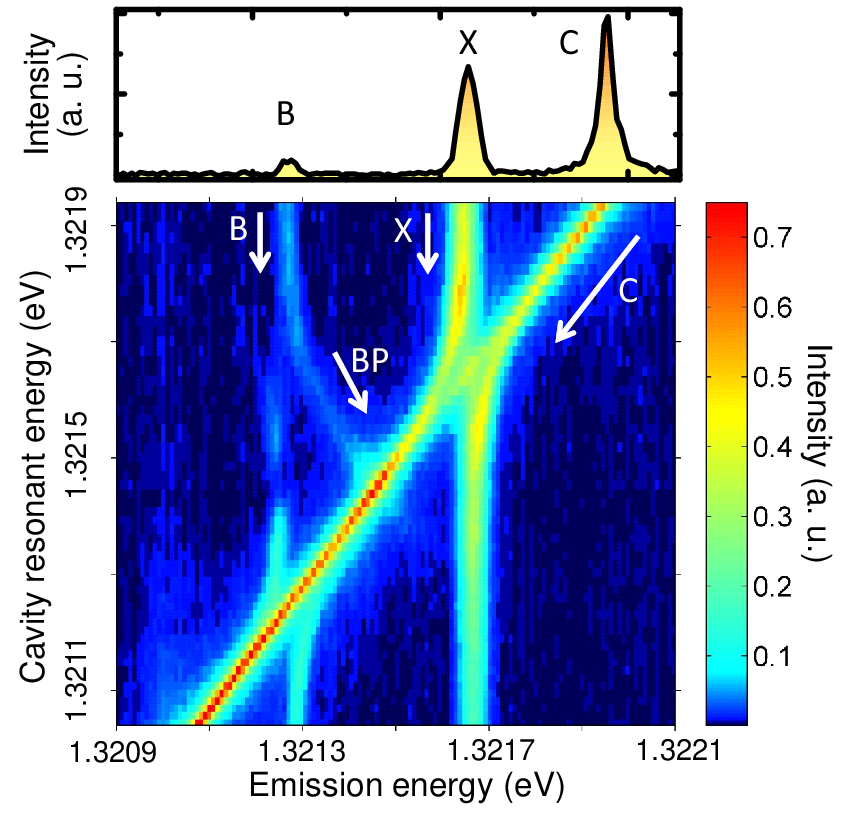}
\caption{(Color online) (Top) Emission spectra under the cavity detuning of $\delta_{c} = \omega_{c}-\omega_{X}$ $ \sim$300 $\mu$eV. (Bottom) Colorplot of the measured spectra under various cavity detunings. Whole spectra are normalized to the highest peak in the series of the cavity scanned spectra.}
\label{fig:f2}
\end{figure}

The top panel of Fig.~\ref{fig:f2} shows an emission spectrum from the QD-cavity system under a detuning $\delta_{c} = \omega_{c}-\omega_{X}$ of $ \sim$300 $\mu$eV. The three peaks (from left to right) are the biexciton to exciton transition (denoted by B), the exciton to ground state transitions (X) and the cavity mode emission (C). The observed cavity quality factor is about 55,000 ($\kappa = 24 \mu eV$). The biexciton binding energy $\chi$ was $\sim$ 400 $\mu$eV. The QD emission lines are broader than that of the cavity presumably due to spectral diffusion and unresolved fine structure splitting.

A colorplot of a series of emission spectra under various QD-cavity detunings is shown in the bottom panel of Fig.~\ref{fig:f2}. The cavity mode frequency was controlled by a xenon gas deposition technique~\cite{Mosor2005}. When the cavity approaches X transition lines, clear anticrossing behavior was observed, demonstrating that the system is in the strong coupling regime. At the resonance condition $\omega_{c} = \omega_{X} $ = 1.32166 eV, we observed a vacuum Rabi splitting of 102 $\mu$eV. The resulting coupling parameter $g_{X}$ is 51 $\mu$eV and thus the ratio $g_{X}/\kappa$ becomes 2.1. The measured Rabi spectrum contains three peaks as observed and explained in several publications~\cite{Hennessy2007,Winger2008a,Yamaguchi2009a,Ota2009b}. Simultaneously, doublet emission is clearly observed around the B line. The two peaks are a result of decay of $\ket{B}$ to the C-X polariton doublet states as reported by Winger \textit{et al.}~\cite{Winger2008a}. We denote one of the polariton line as BP, which directs toward the crossing point with the C line.

The cavity also shows strong coupling with the B line with the spectral triplet feature. The observed vacuum Rabi splitting, under the resonant condition $\omega_{c} = \omega_{X} -\chi $ = 1.32127 eV, was 85 $\mu$eV, which is similar to the X-C case, and the QD-cavity coupling constant for the B polariton $g_{B} \sim$ 43 $\mu$eV is deduced. The Rabi doublet around the B line is formed through $\ket{B,0} \pm \ket{X,1} \rightarrow \ket{X,0}$ transitions via cavity photon leakage. In principle, the B-C polariton can also decay to $\ket{G,1}$ by emitting a single photon to free space mode and the transition is supposed to form doublet emission around the X line. However, the expected two lines are not visible in the experimental result. This is because the decay of B-C polariton is dominated by the cavity photon leakage ($\kappa /2\pi$= 5.8 GHz) and the decay to $\ket{G,1}$ occurs much less frequently. Note that, in this case, the single photon decay rate $\gamma_{B,X}$ would be slower than the intrinsic QD lifetime ($\sim$ 1 GHz) due to photonic bandgap effect and to suppressed Purcel effect under the large cavity detuning of $\sim$ 400 $\mu$eV.

Now we discuss interesting features in the emission spectra around the two photon resonant condition $\omega_{c} = \omega_{X} -\chi /2$ = 1.32147 eV. First, one can see emission enhancement from the cavity mode around the resonance, as predicted in the literature~\cite{DelValle2010, Hohenester2010}. At the same time, strong and weak reduction of the B and X emission, respectively, are also observed (See also Fig.~\ref{fig:f4}). The reduction of the B emission can occur when additional paths open for the decay of $\ket{B,0}$ other than to $\ket{X,0}$ and $\ket{Y,0}$ at the rate $\gamma _{BX}+\gamma _{BY}$. The additional path is the two photon transition to $\ket{G,2}$, at the rate $\Gamma$, and then to $\ket{G,0}$ via cavity photon leakage. With a simple consideration on the transition rates, we can deduce that one biexciton converted into $\frac{\Gamma}{\Gamma+\gamma_{BX}+\gamma_{BY}}$ pairs of two photons. As the reduction of the B line is significant and the relating single photon processes strongly suppressed, $\Gamma$ is expected to be much more than $\gamma_{BX}+\gamma_{BY}$, which is to say, the efficient conversion of the biexciton into the two photon state occurs. The fast two photon decay also causes the reduction of X emission, because the supply of population to $\ket{X,0}$ from $\ket{B,0}$ decreases and the X emission is driven only by carriers incoherently pumped from $\ket{G,0}$. Successful observation of the TPE relies largely on advantageous properties of the QD-PhC cavity system. The strong field in the cavity dramatically enhances the two photon process and the PhC bandgap effect suppress other competing single photon processes. Moderately large $\chi$ gives sufficient detuning for reducing unwanted Purcell effect for the decay of $\ket{B}$ to $\ket{X}$. Note that the BP line is also affected by the two photon resonance around the crossing point with C line and shows emission enhancement there. 

\begin{figure}
\centering
\includegraphics[width=7cm]{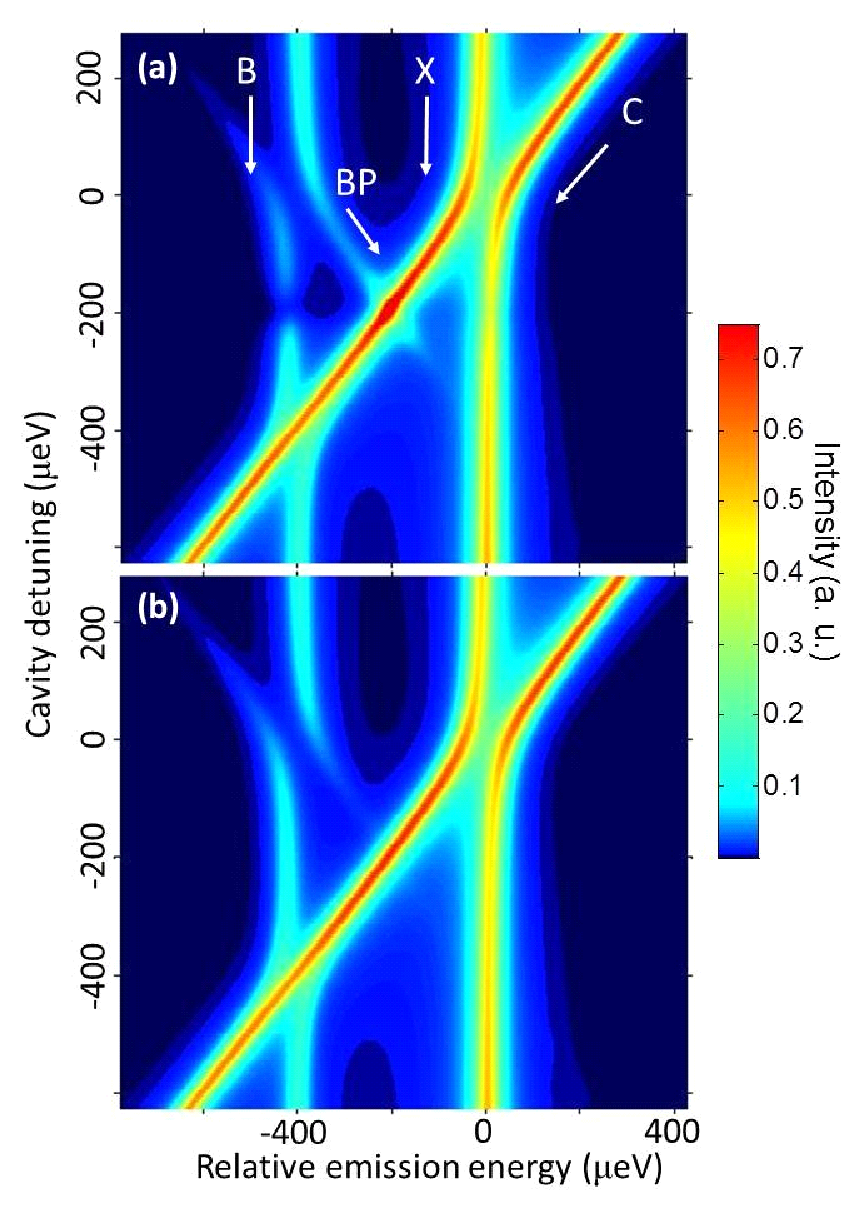}
\caption{(Color online) Colorplot of the calculated spectra under various cavity detunings. The included cavity photon number is up to 2 for (a) and up to 1 for (b). Each spectrum at different detuning conditions in (a) is normalized to the highest peak among the whole spectra. The same normalization factor as (a) is used for (b).}
\label{fig:f3}
\end{figure}

\begin{figure}
\centering
\includegraphics[width=\linewidth]{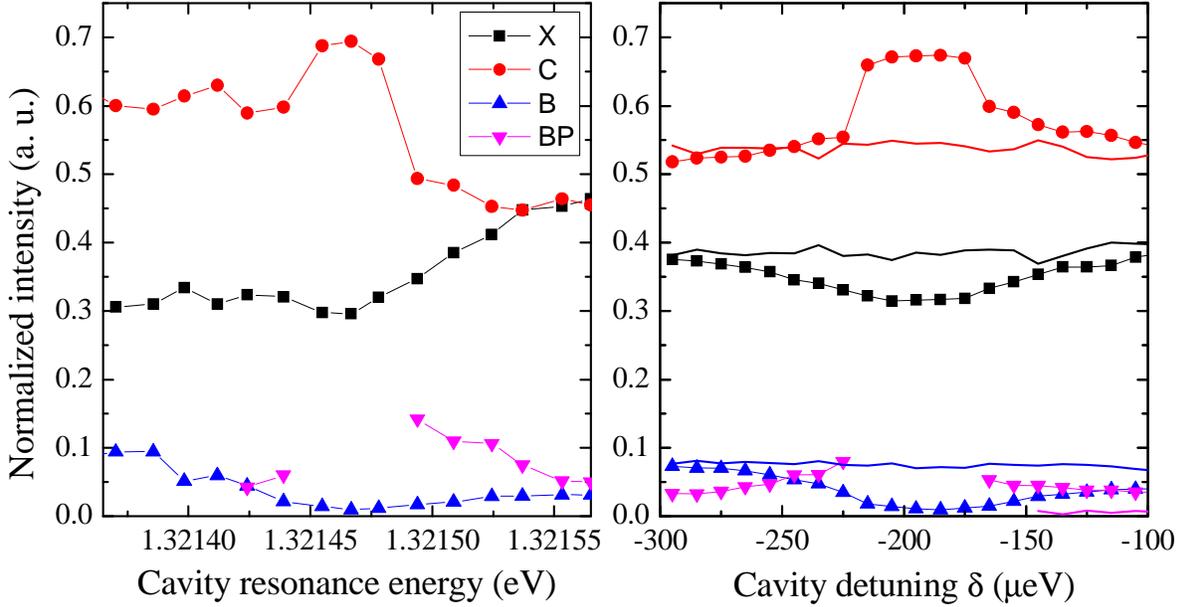}
\caption{(Color online) Integrated intensities of the several emission lines around the two photon resonance condition.  (a) Experimental and (b) calculation results corresponding to Fig~\ref{fig:f3}. The data is normalized to the sum of the whole emission intensity at each detuning. X, C, B and BP lines are expressed as black square, red ball, blue up triangle and magenta down triangle, respectively. Solid lines in (b) is from the calculation including the limited photon number up to one.}
\label{fig:f4}
\end{figure}

To further investigate the observed TPE, we performed numerical calculations based on a quantum master equation including the photon number state up to 2. The model consists a QD described as the 4 level in Fig.~\ref{fig:f1}, coupled with a single cavity mode. The model is based on one used in ref.~\cite{DelValle2010}. The parameters for the calculations were taken from fitting to the experimentally obtained PL spectra and details of the calculation can be found in the Supplemental information. Sum of the emission spectra from the cavity and QD are calculated under incoherent pumping between the dipole allowed transitions in the QD. The calculated results are shown in Fig.\ref{fig:f3} (a). There is good agreement between the experimental results in Fig.\ref{fig:f2} and the theoretical calculations. A noticeable difference exists in appearance of the additional peak between the X-C vacuum Rabi doublet. Inclusion of small contributions from background oscillators might explain the difference. However, these are extraneous effects and do not affect on our conclusion. Note that calculations including the photon number more than 2 shows almost the same results with that includes up to 2. We also calculated the spectra excluding the contributions from two photon processes by truncating the photon number states higher than 1. The calculation result is plotted in Fig.~\ref{fig:f3} (b). The plot does not show any peak in the C line and dip for the B and X lines around the two photon resonance. The BP line in this calculation also shows a monotonic reduction. Through those comparisons, we concluded observed peaks and dips in emission lines in the experiment is a strong evidence for occurrence of the TPE. The calculation also reveals that the photon number in the cavity mode is on the order of 10$^{-2}$ throughout the cavity scan. The small photon number much below one indicates that the observed TPE is spontaneous.

In order to quantitatively discuss the data, we extracted peak intensities of the emission spectra by multiple peak fitting in the detuning range $\delta_{c}$ from -100 to -300 $\mu$eV and plot them in Fig.\ref{fig:f4}. The calculated spectra including the limited photon number up to one are also fitted and expressed as solid lines. Lorentzian curve convolved with spectrometer (Gaussian) is used for the fitting peak function. Overall behaviors of the extracted integrated intensities for the experiment show good agreement with that for the calculation. For the both plots, concerted increase and decrease between C, X, B and BP emissions can be seen and the cavity contribution at the two photon resonance becomes up to 70 $\%$ of the whole emission intensity.  Supposing there is no two photon resonance (see the solid lines), the contribution of the cavity at this points is estimated to be roughly 60 $\%$. Thus, two photon process is considered to be contributing about 10 $\%$ of the whole emission at the two photon resonance. The fraction corresponds with the sum of contributions from the B and BP lines out of any resonances. This coincidence indicates that carriers in $\ket{B}$, which give rise to the B and BP lines via single photon emission, are mostly contributing to C by TPE at its resonance. For converting more carriers into the two photon state, the population in $\ket{B}$ should be increased. If we omit other irrelevant charged exciton states, the dominant path supplying carriers to $\ket{B}$ is supposed to go via $\ket{Y}$ which does not couple to the cavity. The slow decay rate $\gamma_{B,Y}$ and $\gamma_{Y,G}$ due to the photonic bandgap effect is suitable for conveying carriers to $\ket{B}$ without significant loss. Both in the experiment and calculations, we did not specifically select the pumping channel. If we could selectively pump the path via $\ket{Y}$ (e.g. resonant laser excitation with the selected polarization), it would be possible to induce TPE with high efficiency and purity, and largely suppress other single photon decay processes.

The largest discrepancy between the two peak intensity plots in Fig.\ref{fig:f4} (a) and (b) lies in the behavior of the cavity and X branch versus the cavity resonance energy. The calculation results behave almost symmetrically, while the experimental results shows asymmetry across the two photon resonance. Also, X line in the experiment displays an anti-correlation with the cavity intensity and decreases as the cavity goes lower frequency. We consider that these behaviors arise from the phonon mediated coupling process as discussed in the literature~\cite{Hohenester2010,Hohenester2009,*Ota2009a,*Tarel2010,*Hughes2010}, which demonstrates stronger cavity emission as the detuning moves towards the red. Future studies will reveal the effect of phonons on such exciton-biexciton-cavity coupling systems. 

In summary, we demonstrated spontaneous TPE from a single QD placed inside a high Q PhC nanocavity. We detected the TPE process in changes of the PL spectra from the cavity, biexciton and exciton states of the QD, depending on the cavity detunings. Calculations including the two photon contribution could well reproduce the observed spectra, while another calculation excluding the two photon contribution could not; thus validating the occurrence of the TPE. The demonstrated strong optical nonlinearity in a QD-PhC nanocavity system will pave a way to quantum information devices using two photon processes in QDs~\cite{Lin2010,*Majumdar2010b} and to QD-based two photon lasers~\cite{DelValle2010}, and will give a significant insight into studies of few-photon optical nonlinearity in QD-cavity systems~\cite{Englund2007,*Srinivasan2007,*Bose2011}.

\begin{acknowledgments}
We thank S. Ishida, S. Ohkouchi, M. Shirane, Y. Igarashi, M. Nomura, R. Ohta, E. Harbord and K. Watanabe for their technical support and for fruitful discussions. This research was supported by the Special Coordination Funds for Promoting Science and Technology, Japan.
\end{acknowledgments}

\begin{flushleft} 
\textit{\textbf{Supplementary material - Calculation}}
\end{flushleft}

We calculated the emission spectra from the biexciton/exciton-cavity coupled system using a master equation based method. An overview of the model has already been discussed in the main text. The system for the calculation is spun in the Hilbert space by $\ket{i,n_{c}} _{i=B,X,Y,G}$, where $n _{c} = 0,1,2...$ is the cavity photon number. The upper limit of $n _{c}$ is set to 2 by truncating the states with higher $n _{c}$. This approximation is enough to describe our case with low $n _{c} \ll 1$ in all the situation. The system Hamiltonian $H$ in the Sch\"{o}dinger picture under the rotating wave and dipole approximation is given by

\begin{equation}
H=\hbar \sum^{}_{i=B,X,Y,G} \omega_{i}\sigma_{ii} +\hbar \omega_c a^{\dagger}a+ \hbar g _{X}(\sigma_{GX}a^{\dagger} +H.c. ) + \hbar g _{B}(\sigma_{XB}a^{\dagger} +H.c. )
\label{eq:tpe1}
\end{equation}

, where $\sigma _{ij}=\ket{i}\bra{j}$ is the pseudo Pauli spin operator, $\omega_{i}$ and $\omega_c$ are the eigenfrequencies of the state $\ket{i}$ and the cavity mode, respectively. $a$ is the annihilation operator of the cavity mode. The cavity is coupled with $\ket{G}\leftrightarrow\ket{X}$ and  $\ket{X}\leftrightarrow\ket{G}$ transitions with the single photon Rabi frequency of $g_{X}$ and $g_{B}$, respectively. 

We set the $\omega_{G}=0$ as the zero point of energy and take rotating frame at the $\ket{X} \rightarrow \ket{G}$ transition frequency $\omega_{X}$. Then obtain,
  
\begin{equation}
H'=\hbar (-\chi \sigma_{BB}+\delta \sigma_{YY}) +\hbar \delta_c a^{\dagger}a+ \hbar g _{X}(\sigma_{GX}a^{\dagger} +H.c. ) + \hbar g _{B}(\sigma_{XB}a^{\dagger} +H.c. )
\label{eq:tpe2}
\end{equation}

, where $ \chi = -\omega_{B} +2\omega_{X}$ , $\delta = \omega_{Y}-\omega_{X}$,  $\delta_{c} = \omega_{c}-\omega_{X}$. $\chi$ is the biexciton binding energy. $\delta$ accounts for the fine structure splitting between the linearly polarized two excition states. $\delta_{c}$ is the cavity detuning from the $\ket{X}\leftrightarrow\ket{G}$ exciton line. The dynamics of the system can be calculated using the master equation under Born-Markov approximation, 

\begin{equation}
\frac{\partial \rho}{\partial t} = -\frac{i}{\hbar}[H',\rho]+L\rho .
\label{eq:tpe3}
\end{equation}

Here, we incorporated Markovian processes $L\rho$, which is given by 

\begin{eqnarray} 
L\rho = \frac{\kappa}{2} \mathcal{L}_{decay}(a)\rho + \frac{\gamma}{2} \{ \mathcal{L}_{decay}(\sigma _{XB})\rho +\mathcal{L}_{decay}(\sigma _{YB})\rho +\mathcal{L}_{decay}(\sigma _{GX})\rho +\mathcal{L}_{decay}(\sigma _{GY})\rho \} \nonumber \\
+ \frac{P}{2} \{ \mathcal{L}_{pump}(\sigma _{XB})\rho + \mathcal{L}_{pump}(\sigma _{YB})\rho +\mathcal{L}_{pump}(\sigma _{GX})\rho +\mathcal{L}_{pump}(\sigma _{GY})\rho \} \nonumber \\
+ \frac{\gamma _{phase}}{2} \{ \mathcal{L}_{phase}(\sigma_{BB} ,\sigma_{XX} )\rho +\mathcal{L}_{phase}(\sigma_{BB} ,\sigma_{YY})\rho +\mathcal{L}_{phase}(\sigma_{XX} ,\sigma_{GG})\rho +\mathcal{L}_{phase}(\sigma_{YY} ,\sigma_{GG})\rho \}
\label{eq:tpe4}
\end{eqnarray}

Here, we define functions $\mathcal{L}_{decay}(x)$,$\mathcal{L}_{pump}(x)$,$\mathcal{L}_{phase}(x,y)$ by following reference~\cite{DelValle2010}. 
   
\begin{eqnarray}
\mathcal{L}_{decay}(x)\rho &=& 2x^{\dagger}\rho x - x^{\dagger}x\rho -\rho x^{\dagger}x ,\nonumber \\
\mathcal{L}_{pump}(x)\rho &=& 2x\rho x^{\dagger} - x x^{\dagger}\rho -\rho x x^{\dagger} , \\
\mathcal{L}_{phase}(x,y)\rho &=&  (x^{\dagger}x-y^{\dagger}y)\rho   (x^{\dagger}x-y^{\dagger}y) -\rho \nonumber
\label{eq:tpe5}
\end{eqnarray}

Here, $\kappa, \gamma, P, \gamma_{phase}$ respectively correspond to the cavity loss, decay via spontaneous emission, incoherent pumping between the QD levels, and pure dephasing of QD's levels. Temperature dependent terms in the non-Markovian processes are omitted. This is valid in our case because the system is operated in the optical regime at the low temperature of $\sim$ 4.5 K. We calculated the steady state emission spectra from the system under various $\delta_{c}$. First we computed the steady state of the system using the equation~\ref{eq:tpe3} by setting $\frac{\partial}{\partial t} = 0$. Then, two time correlation functions, which are necessary to obtain the spectra, are calculated via the quantum regression theorem. From the correlation functions, the emission spectra from the each leakage path were calculated by Wiener-Khintchine theorem. The sum of them at a $\delta_{c}$ becomes a spectrum we plotted in this work. These calculation of the emission spectra was aided by Quantum Optics Tool Box~\cite{Tan1999}. 

The parameters $g_{X}$ = 51 $\mu$eV, $g_{B}$ = 43 $\mu$eV, $\kappa = 24  \mu \mathrm{eV}$ and $\chi = 400  \mu \mathrm{eV}$ are extracted from the experimental data by multiple Lorentzian peak functions convolved with a Gaussian peak function (FWHM 23 $\mu$eV). We also provide finite values for other Markovian proceses, $\gamma = 0.13 \mu \mathrm{eV}$, $P = 0.05 \mu \mathrm{eV}$, and $\gamma_{phase} =  5 \mu \mathrm{eV}$. $\gamma/2 \pi \sim  0.2  \mathrm{GHz} $ is a typical value for spontaneous emission rate of QDs in a two dimensional photonic bandgap. Although we could not resolve $\delta$ by our setup, we set it to 10 $\mu$eV. This value is typical for our QDs at the wavelength of $\sim$ 940 nm. Although the transitions related to $\ket{Y}$ are not coupled to the cavity mode, the existence of the state $\ket{Y}$ is important for efficiently pumping carriers to the state $\ket{B}$. The value of $P$ is determined so that the calculation spectra are quantitatively muches with the experimental data. Calculations with different order of magnitude of $P$ cannot account for the experimental results. For revealing the effect of the two photon processes, we changed upper limit of the $n_{c}$ from 4 to 1 by truncating the states with higher photon number.

\bibliography{MyCollection}

\bibliographystyle{apsrev4-1}

\end{document}